\documentclass[11pt]{article}

\usepackage[english]{babel}
\usepackage[latin1]{inputenc}
\usepackage[T1]{fontenc}
\usepackage[usenames]{color}
\usepackage{amsmath}
\usepackage{amssymb}
\usepackage{hhline}
\usepackage{multirow}
\usepackage{colortbl}
\usepackage{enumerate}
\usepackage{array}
\usepackage{soul}
\usepackage{graphicx}
\usepackage{eurosym}
\usepackage{mathrsfs}
\usepackage{dsfont}
\usepackage{amsthm}
\usepackage{makecell}
\usepackage[titletoc]{appendix}
\usepackage[titles]{tocloft}
\usepackage{indentfirst}

\newtheorem{lemma}{Lemma}[section]
\newtheorem{definition}{Definition}[section]

\title{Chebyshev Reduced Basis Function applied to Option Valuation. \thanks{Research supported by Spanish MINECO
under grants MTM2013-42538-P and MTM2016-78995-P. Both authors acknowledge the helpful comments of Mich\`{e}le Breton and Peter Christoffersen.}}

\author{Javier de Frutos\thanks{Instituto de Matem\'{a}ticas (IMUVA), Universidad de Valladolid, Paseo de Bel\'{e}n 7, Valladolid, Spain. e-mail:frutos@mac.uva.es} and V\'{\i}ctor Gat\'{o}n\thanks{Instituto de Matem\'{a}ticas (IMUVA), Universidad de Valladolid,  Paseo de Bel\'{e}n 7, Valladolid, Spain. e-mail:vgaton@mac.uva.es}}

\begin{document}

\maketitle

\begin{abstract}
We present a numerical method for the frequent pricing of financial derivatives that depends on a large number of variables. The method is based on the construction of a polynomial basis to interpolate the value function of the problem by means of a hierarchical orthogonalization process that allows to reduce the number of degrees of freedom needed to have an accurate representation of the value function. In the paper we consider, as an example,
a GARCH model that depends on eight parameters and show that a very large number of contracts for different maturities and asset and parameters values can be valued in a small computational time with the proposed procedure.
In particular the method is applied to the problem of model calibration. The method is easily generalizable to be used with other models or problems.

\end{abstract}

\section{Introduction}

This paper concerns the design of a Reduced basis function approach to mitigate the impact of the ``Curse of Dimensionality'' which appears when we deal with multidimensional interpolation, in particular, when we price financial derivatives employing multivariable models, like GARCH, or whose price may depend on multiple assets which follow different stochastic processes.

Since the market prices change almost constantly, thousands of derivative prices have to be recomputed very fast, hence numerical techniques that allow fast evaluation of multidimensional models are of interest.

Several techniques can be found in the literature to solve multidimensional option pricing problems. For example, Monte Carlo methods are very popular since they are straightforward to implement and handle multiple dimensions, although their convergence rates tend to be slow. Several approaches, like variance reduction techniques can be employed to increase their velocity (see \cite{Giles} or \cite{Glasserman}). Lately, PDE methods have also been employed in financial problems (see, for example, \cite{Breton}, \cite{FrutosGaton1}, \cite{Linde} or \cite{Pironneau}) due to their faster speed of convergence with respect to Monte Carlo methods. In \cite{Bungartz}, an adaptive sparse grid algorithm using the finite element method is employed for the solution of the Black-Scholes equation for option pricing. A Legendre series expansion of the density function is employed in \cite{Hok} for option valuation. In \cite{Balajewicz}, a proper orthogonal decomposition and non negative matrix factorization is employed to make pricing much faster within a given model parameter variation range. A general review of financial problems or models, numerical techniques and software tools can be found in \cite{DuanHandbook}.

The main objective of this work is to reduce the computing time and storing costs which may appear in multidimensional models. The proposed method has two differentiated steps. In the first one (off-line computation), an approximation to the option pricing function is computed through a polynomial Reduced Basis method. This step usually is expensive in terms of computational cost, but it is performed only once. In the second step (on-line computation), we employ the polynomial constructed in the previous step to price a large number of contracts or calibrate model parameters. In this second step, the big gain in computational time arises because, as the polynomial does not need to be recomputed, it can be used as many times as needed.

One way to compute numerical approximations of the value of multidimensional functions is to compute the function value in a given set of nodes and to use polynomial interpolation. Unfortunately, for a high number of dimensions, the memory requirements to store the interpolant and  the computational time required for each evaluation
grow exponentially. This effect is known as the ``Curse of Dimensionality''.

A common approach to reduce the curse of dimensionality is to search the dimensions where increasing the number of  interpolation nodes would reduce faster the interpolation error. The  method that we propose here approaches, someway, the problem from the other side. Instead on focusing on the construction of the interpolation polynomial, what we propose is to obtain, from the interpolant, a new reduced polynomial which gives similar accuracy as the original one but requiring much less computational effort.

In order to fix ideas we present the method with Chebyshev interpolation although the techniques in this paper are easily generalizable. The properties of Chebyshev polynomials, \cite{Rivlin}, enable us to use time-competitive and accurate
fast fourier transform methods for computing the coefficients, evaluation and differentiation of the polynomials.

In first place, we show how to build the interpolant and we design a very efficient evaluation algorithm which allows to compute the polynomial value for different values of each of the parameters simultaneously, something that will be referred as {tensorial evaluation}. Afterwards, for a fixed interpolant, we propose a reduced basis approach employing a Hierarchical orthonormalization procedure along each one of the dimensions.
This procedure rewrites the interpolant in function of a set of orthonormal function basis which are ordered hierarchically depending on the amount of information of the interpolant that they posses. Afterwards, fixed a tolerance level for the error with respect to the interpolant, we retain the minimum number of functions in the basis such that this tolerance is fulfilled. Furthermore, since the function basis are written in function of Chebyshev polynomials, the evaluation algorithm previously designed can be adapted to the new approximation. The result is a polynomial which approximates the multidimensional function and requires much less memory capacity and computational time for evaluation.

The paper is organized as follows.  In Section \ref{RBFAOVPi}, a fast tensorial Chebyshev polynomial interpolation is developed. The precision is obtained increasing the number of interpolation nodes, which leads to the storage-cost problem. In Section \ref{RBFAOVRBA} the reduced approximation is presented. Finally, in Section \ref{RBFAOVNR} we perform numerical experiments with the different techniques developed over a multidimensional model employed in pricing financial derivatives.

All the algorithms presented in this work have been implemented in Matlab vR2010a. All the numerical experiments have been realized in a personal computer with an Intel(R) Core(TM) i3 CPU , 540 @ 3.07GHz, memory RAM of 4,00 GB and a 64-bits operative system.

\section{Polynomial Interpolation}\label{RBFAOVPi}

We propose a Chebyshev polynomial interpolation procedure for multidimensional models. The interpolation is done using Chebyshev polynomials and nodes in the intervals where the parameters are defined.

\begin{definition}\label{Ch1defChebypol} Let us define
\begin{equation}
T_n(x)=\cos\left(n \arccos (x) \right),
\end{equation}
where $0\leq \arccos (x) \leq \pi$.
\end{definition}
It is well known, \cite{Rivlin}, that this function is a polynomial of degree $n$, called the Chebyshev polynomial of degree n.
\begin{definition}\label{Ch1defChebynodes} Let $N\in \mathbb{N}$. The $N+1$ Chebyshev nodes $\{\tilde{\alpha}^k\}_{k=0}^{N}$ in interval $[a, \ b]$ correspond to the extrema of $T_n(x)$ and they are given by:
\begin{equation}\label{Ch1forChbnodes}
\tilde{\alpha}^k=\frac{1}{2}\left[\cos\left(\frac{\pi k}{N}\right)(b-a)+(b+a)\right], \quad k=0,1,...,N.
\end{equation}

We also define the $N+1$ Chebyshev nodes $\{\alpha^k\}_{k=0}^{N}$ in interval $[-1,1]$, where
\begin{equation}
\alpha^k=\cos\left(\frac{\pi k}{N}\right), \quad k=0,1,...,N.
\end{equation}
\end{definition}

Here we present just the definitions that are needed for the proposed method.  The practical computation of the coefficients of the interpolant is postponed to Subsection \ref{Ch1COTIP}.

\begin{definition}\label{Ch1defpolinternvariable}
Let $\tilde{\textbf{x}}=\left(\tilde{x}_1,\tilde{x}_2,...,\tilde{x}_n \right)$ and $\tilde{F}(\tilde{\textbf{x}})$ be a continuous function defined in $\tilde{x}_j\in[\tilde{x}^{\min}_j, \ \tilde{x}^{\max}_j], \quad j=1,2,...,n$.

For $\textbf{x}=\left(x_1,x_2,...,x_n \right)$, we define the function $F(\textbf{x}), \ \textbf{x}\in[-1,1]^n$ as
\begin{equation*}
F(\textbf{x})=\tilde{F}(\tilde{\textbf{x}}),
\end{equation*}
where
\begin{equation}\label{Ch1camcoordmult}
\tilde{x}_j=\frac{\tilde{x}^{\max}_j-\tilde{x}^{\min}_j}{2}x_j+\frac{\tilde{x}^{\max}_j+\tilde{x}^{\min}_j}{2}, \quad
\begin{aligned}
& x_j\in[-1, \ 1], \\
& j=1,2,...,n.
\end{aligned}
\end{equation}

For $\boldsymbol{N}=\{N_1,N_2,...,N_n\}\in\mathbb{N}^n$, we define
\begin{equation}
L^{\boldsymbol{N}}=\left\{\boldsymbol{l}=(l_1,l_2,...,l_n) \ , \ 0\leq l_j\leq N_j, \ j=1,2,...,n\right\}.
\end{equation}

For $j=1,2,...,n$, let $\left\{\alpha^k_j\right\}_{k=0}^{N_j}$ be the $N_j+1$ Chebyshev nodes in $[-1, \ 1]$ and $\left\{\tilde{\alpha}^k_j\right\}_{k=0}^{N_j}$ be the  $N_j+1$ Chebyshev nodes in $[\tilde{x}^{\min}_j, \ \tilde{x}^{\max}_j]$.

We use the notation $\boldsymbol{\alpha}^{\boldsymbol{l}}= \left(\alpha^{l_1}_1, \ \alpha^{l_2}_2,...,\alpha^{l_n}_n \right)$ and $\tilde{\boldsymbol{\alpha}}^{\boldsymbol{l}}=\left(\tilde{\alpha}^{l_1}_1,\tilde{\alpha}^{l_2}_2,...,\tilde{\alpha}^{l_n}_n \right)$.

Let $I_{\boldsymbol{N}} F(\textbf{x})$ be the n-dimensional interpolant of function $F(\textbf{x})$ at the Chebyshev nodes $\left\{\boldsymbol{\alpha}^{\boldsymbol{l}}\right\}_{\boldsymbol{l}\in L^{\boldsymbol{N}}}$, i.e. the polynomial which satisfies
\begin{equation*}
I_{\boldsymbol{N}} F(\boldsymbol{\alpha}^{\boldsymbol{l}})=F(\boldsymbol{\alpha}^{\boldsymbol{l}})=\tilde{F}(\tilde{\boldsymbol{\alpha}}^{\boldsymbol{l}}), \quad \boldsymbol{l}\in L^{\boldsymbol{N}}.
\end{equation*}

Polynomial $I_{\boldsymbol{N}} F(\textbf{x})$ is given by
\begin{equation}
I_{\boldsymbol{N}} F(\textbf{x})=\sum_{\boldsymbol{l}\in L^{\boldsymbol{N}}}\hat{p}_{\boldsymbol{l}}T^{\boldsymbol{l}}(\textbf{x}), \quad \textbf{x}\in[-1, 1]^{n},
\end{equation}
where
\begin{equation*}
\begin{aligned}
\hat{p}_{\boldsymbol{l}} & =\hat{p}_{(l_1,l_2,...,l_n)}\in \mathbb{R}, \\
T^{\boldsymbol{l}}(\textbf{x}) & =T_{l_1}(x_1)T_{l_2}(x_2) ... T_{l_n}(x_n).
\end{aligned}
\end{equation*}
\end{definition}

Let $\tilde{\Omega}=\prod_{j=1}^n [\tilde{x}^{\min}_j, \ \tilde{x}^{\max}_j]$ and suppose that we need a numerical approximation for several different values in each of the variables. Let $\boldsymbol{\Theta}\in\tilde{\Omega}$ be a set of values such that for $q_j\in\mathbb{N}, \ 1 \leq j \leq n $,
\begin{equation} \label{defconjtheta}
\boldsymbol{\Theta}=\left\{\tilde{\boldsymbol{x}}=(\tilde{x}_1,\tilde{x}_2,...,\tilde{x}_n)  /  \tilde{x}_j \in \{\tilde{x}^1_j,...,\tilde{x}^{q_j}_j\}, \ \tilde{x}^k_j\in [\tilde{x}^{\min}_j, \ \tilde{x}^{\max}_j], \ 1\leq k \leq q_j \right\}
\end{equation}
where we note that the number of points in $\boldsymbol{\Theta}$ is $|\boldsymbol{\Theta}|=\prod_{j=1}^n q_j$.

The numerical approximation is computed with the polynomial $I_{\boldsymbol{N}} F(\textbf{x})$, where the relation between $\tilde{\boldsymbol{x}}$ and $\boldsymbol{x}$ is given by formula (\ref{Ch1camcoordmult}). The algorithms presented in Subsection \ref{Ch1TEDIP} allow us to evaluate the interpolation polynomial in a set of points like $\boldsymbol{\Theta}$ very fast, something that from now on will be referred as {tensorial evaluation}. This is achieved through a suitable defined multidimensional array operation and the employment of efficient algorithms described in Subsection \ref{Ch1TEDIP}.

\subsection{Computation of the interpolating polynomial.} \label{Ch1COTIP}

\noindent \textbf{Univariate case}

Let $\tilde{F}(\tilde{x})$ be a continuous function defined in $\tilde{x}\in[\tilde{x}^{\min}, \ \tilde{x}^{\max}]$ and suppose that we want to compute the Chebyshev interpolant
\begin{equation*}
I_{N_1}F(x)=\sum^{N_1}_{l=0}\hat{p}_{l}T_l(x), \quad x\in[-1, 1].
\end{equation*}

It must hold that
\begin{equation*}
\begin{aligned}
F({\alpha}^k)&=\sum^{N}_{l=0}\hat{p}_{l}T_l(\alpha^k)=\sum^{N}_{l=0}\hat{p}_{l}\cos(l(\arccos(\alpha^k)))\\
&=\sum^{N}_{l=0}\hat{p}_{l}\cos\left(l\frac{\pi k}{N}\right),
\end{aligned}
\end{equation*}
where $\{\tilde{\alpha}^k\}^{N}_{k=0}, \ \{\alpha^k\}^{N}_{k=0}$ are the Chebyshev nodes in $[\tilde{x}^{\min}, \tilde{x}^{\max}]$ and $[-1, \ 1]$.

There are several efficient algorithms that allow us to obtain the coefficients $\{\hat{p}_{l}\}^N_{l=0}$. For the univariate case, we employ the algorithm presented in \cite{Canuto}.

\noindent \textit{Algorithm C1v:}

1. Construct
\begin{equation*}
z=\left[F(\alpha^0), F(\alpha^1),..., F(\alpha^{N-1}), F(\alpha^{N}), F(\alpha^{N-1}), ...,F(\alpha^2), F(\alpha^1)\right]^T.
\end{equation*}

2. Compute
\begin{equation*}
y=\frac{\text{real}\left(\text{FFT}(z)\right)}{2N}.
\end{equation*}

3.
\begin{equation*}
\left\{
\begin{aligned}
& \hat{p}_{0}=y(1), \\
& \hat{p}_{l}=y(l+1)+y(2N-(l-1)) \ \text{if} \  0<l<N, \\
& \hat{p}_{N}=y(N). \\
\end{aligned}
\right.
\end{equation*}

\noindent \textbf{Multivariate case}

Let $\tilde{F}(\tilde{\textbf{x}})$ and $\textbf{N}\in\mathbb{N}^n$ be as given in Definition \ref{Ch1defpolinternvariable} and suppose that we want to construct the interpolant
\begin{equation*}
I_{\boldsymbol{N}} F(\textbf{x})=\sum_{\boldsymbol{l}\in L^{\boldsymbol{N}}}\hat{p}_{\boldsymbol{l}}T^{\boldsymbol{l}}(\textbf{x}), \quad \textbf{x}\in[-1, 1]^{n}.
\end{equation*}

\begin{definition}\label{Ch1defpermutation}
Let $A$ be an array of dimension $n_{1}\times n_{2}\times ... \times n_{m}$. We denote the vector
\begin{equation*}
A(j_1,...,j_{n_{s-1}},:,j_{n_{s+1}},...,j_m)=\{A(j_1,...,j_{n_{s-1}},j,j_{n_{s+1}},...,j_m)\}^{n_s}_{j=1},
\end{equation*}
where $1\leq j_i\leq n_i, \quad \forall i\in\{1,2,...,m\}-\{s\}$.

Let $B$ be an array of dimension $a\times n_{1}\times n_{2}\times ... \times n_{m}$. We define the permutation operator $\mathcal{P}$ such that if:
\begin{equation*}
D=\mathcal{P}(B),
\end{equation*}
we have that $\dim(D)=n_{1}\times n_{2}\times ... \times n_{m} \times a$ and
\begin{equation*}
D(j_1,...,j_m,:)=B(:,j_1,...,j_m).
\end{equation*}
\end{definition}

Suppose that we have already computed the function value at the Chebyshev nodes, i.e., $\forall k_j  \in \{0,...,N_j\}, \ j=1,2,...,n,$
\begin{equation*}
\tilde{F}\left(\tilde{\alpha}^{k_1}_1,\tilde{\alpha}^{k_2}_2,...,\tilde{\alpha}^{k_n}_n\right)=F\left(\alpha^{k_1}_1,\alpha^{k_2}_2,...,\alpha^{k_n}_n\right), \end{equation*}
which are stored in an array $\Gamma_{(N_1+1)\times(N_2+1)\times...\times(N_n+1)}$ such that
\begin{equation*}
\Gamma(k_1+1,k_2+1,...,k_n+1)=F\left(\alpha^{k_1}_1,\alpha^{k_2}_2,...,\alpha^{k_n}_n\right).
\end{equation*}

The coefficients $\hat{p}_{\boldsymbol{l}}$ of the interpolant are obtained through the following algorithm.

\noindent \textit{Algorithm Cnv:}

1. $B_1=\Gamma$.

2. For i=1 to $n$

\hspace{0.5 cm} 2.1. $\{m_1,m_2, ...,m_n\}=\dim(B_i)$.

\hspace{0.5 cm} 2.2. For $j_2=1$ to $m_2$, for $j_3=1$ to $m_3$, ...,  for $j_n=1$ to $m_n$
\begin{equation*}
C_i(:,j_2,j_3,...,j_n)=\text{Algorithm C1v}\left(B_i(:,j_2,j_3,...,j_n)\right).
\end{equation*}

\hspace{0.5 cm} 2.3. $B_{i+1}=\mathcal{P}(C_i)$.

3. $\hat{p}_{\boldsymbol{l}}=B_{n+1}(l_1+1,l_2+1,...,l_n+1)$.

\

We remark that the FFT routine in Matlab admits multidimensional evaluation so that, step 2.2 of the previous algorithm can be efficiently computed without using loops.

\subsection{Tensorial Evaluation and Differentiation of the interpolation polynomial.}\label{Ch1TEDIP}

\begin{definition}\label{Ch1defmultidimmatrixproduct}

Let $A$ and $B$ be two arrays, $(A)_{a\times n_{1}\times n_{2}\times ... \times n_{k}}$ and $(B)_{a\times b}$ respectively, and such that $b>1$.

We define the tensorial array operation $C=A\otimes B$, as the array $C$ given by:
\begin{equation}
C(j_1,...,j_k,:)= \mathcal{P}\left(B' A(:,j_1,...,j_k)\right),
\end{equation}
where $B' A(:,j_1,...,j_k)$ is the usual product of matrix times a vector and  $\mathcal{P}$ is the permutation operator introduced in Definition \ref{Ch1defpermutation}.

It is easy to check that $\dim(C)=n_{1}\times n_{2}\times ... \times n_{k}\times b$.

\end{definition}

Concerning the implementation in Matlab of the tensorial array operation,
\begin{equation*}
C=\text{permute}\left(\text{multiprod}(B',A),[2:n \ 1]\right),
\end{equation*}
where \textit{permute} is a standard procedure implemented in Matlab.

The algorithm \textit{multiprod}, implemented by Paolo de Leva and available in Mathworks (see \cite{deLeva}), makes the required tensorial operation simultaneously in all variables in a very efficient way.

Suppose now that we have a polynomial
\begin{equation*}
I_{\boldsymbol{N}} F(\textbf{x})=\sum_{\boldsymbol{l}\in L^{\boldsymbol{N}}}\hat{p}_{\boldsymbol{l}}T^{\boldsymbol{l}}(\textbf{x}) = \sum_{l_1=0}^{N_1}\sum_{l_2=0}^{N_2}...\sum_{l_n=0}^{N_n}\hat{p}_{\boldsymbol{l}}T_{l_1}(x_1)T_{l_2}(x_2)...T_{l_n}(x_n).
\end{equation*}

We want to evaluate the polynomial in a finite set of points $\boldsymbol{\Theta}$, which was defined in (\ref{defconjtheta}).

\

For computational reasons, we impose that $q_j>1, \ j=1,2,...,n.$ By default, \textit{multiprod} algorithm does not recognize arrays of $q_1\times...\times q_{i-1}\times 1 \times q_{i+1} \times...\times q_n$-dimension and collapses to $q_1\times...\times q_{i-1}\times q_{i+1} \times...\times q_n$-dimension. Since $\otimes$ consists of \textit{multiprod} and a \textit{permutation}, if $q_i=1$, a wrong dimension will be permuted in the evaluation algorithm.

The evaluation algorithm has two steps.

\

\noindent \textbf{1. Evaluate the Chebyshev polynomials:}

We use the recurrence property of Chebyshev polynomials:
\begin{equation*}
T_0(x)=1, \quad T_1(x)=x, \quad T_l(x)=2x T_{l-1}(x)-T_{l-2}(x), \quad l=2,3,4,...,
\end{equation*}
that with the number $\boldsymbol{N}$ of interpolation points involved in the option pricing problem works fairly well.

From the definition of $\boldsymbol{\Theta}$ (see (\ref{defconjtheta})), the possible values of each variable are a finite number.
We employ the notation $\eta_j=\{x^k_j\}_{k=1}^{q_j}$ to denote the corresponding values in $\boldsymbol{\Theta}$ after the change of variables (\ref{Ch1camcoordmult}). Using the recurrence property, we compute
\begin{equation*}
\begin{aligned}
\boldsymbol{T}(\eta_1) &= \boldsymbol{T}(\eta_1)_{(N_{1}+1)\times q_1} = \left(T_{l}(x^k_1)\right)_{0\leq l\leq N_1, \ 1\leq k \leq q_1}, \\
\boldsymbol{T}(\eta_2) &= \boldsymbol{T}(\eta_2)_{(N_{2}+1)\times q_2} = \left(T_{l}(x^k_2)\right)_{0\leq l\leq N_2, \ 1\leq k \leq q_2},   \\
& ... \\
\boldsymbol{T}(\eta_n) &= \boldsymbol{T}(\eta_n)_{(N_{n}+1)\times q_n} = \left(T_{l}(x^k_2)\right)_{0\leq l\leq N_n, \ 1\leq k \leq q_n}, \\
\end{aligned}
\end{equation*}
and we store each result in a two dimensional array.

\noindent \textbf{2. Evaluate the rest of the polynomial.}

The evaluation of the polynomial $I_{\boldsymbol{N}} F(\textbf{x})$ for the whole set of points $\boldsymbol{\Theta}$  can be done at once using the tensorial array operation.

The polynomial coefficients are stored in a $(N_1+1)\times(N_2+1)\times...\times(N_n+1)$-dimensional array $A$.
\begin{equation*}
A(l_1+1,l_2+1,...,l_n+1)=\hat{p}_{(l_1,l_2,...,l_n)},
\end{equation*}
and we compute
\begin{equation}
I_{\boldsymbol{N}} F(\boldsymbol{\Theta})=\left(...\left[\left(A\otimes\boldsymbol{T}(\eta_1)\right)\otimes\boldsymbol{T}(\eta_2)\right]...\right)\otimes\boldsymbol{T}(\eta_n),
\end{equation}
where the result will be an $q_1\times q_2 \times ... \times q_n$-dimensional array which contains the evaluation of the interpolant in all the points of set $\boldsymbol{\Theta}$.

We remark that the previous definition must not be seen as a product with the usual properties. The order of the parenthesis has to be strictly followed in order to be consistent with the dimensions.

\

\noindent \textit{Polynomial differentiation:}

Sometimes, we may also need to compute an approximation to the derivative of the multidimensional function. For example, if we want to find the values of the parameters of the model that approximate best to a given a set of data, in the sense of a least square minimization.

Note that if $\tilde{x}\in[a,b]$ and we have interpolated function $\tilde{F}(\tilde{x})$
\begin{equation*}
\tilde{F}(\tilde{x})\approx I_N F(x)=\sum^{N}_{l=0}\hat{p}_{l}T_l(x),
\end{equation*}
where
\begin{equation*}
\tilde{x}=\frac{b-a}{2}x+\frac{b+a}{2}, \quad x\in[-1,1],
\end{equation*}
we can approximate, if function $F$ is regular enough (see \cite{Canuto}),
\begin{equation*}
\tilde{F}'(\tilde{x}) \approx (I_N F(x))'= \frac{2}{b-a}\sum^{N-1}_{l=0}\hat{q}_{l}T_l(x),
\end{equation*}
where (see \cite[(2.4.22)]{Canuto}) for $l=0,1,...,N-1$:
\begin{equation*}
\hat{q}_l=\frac{2}{c_l}+\underset{j+l \ \text{odd}}{\sum_{j=l+1}^{N}} j \hat{p}_j, \quad \text{where} \  c_l=\left\{\begin{aligned} & 2, \quad l=0, \\
& 1, \quad l\geq1.  \end{aligned}\right.
\end{equation*}

This implies that the coefficients of the derivatives of the polynomials need to be computed each time or stored in memory. Both options do not fit with the objective of this work.

If the coefficients are computed each time they are needed, that increases the total time cost. On the other hand, there is a memory storage problem in the polynomial interpolation technique. To store the coefficients of the derivative means to almost double the memory requirements of the method.

For these reasons, we prefer to employ, if possible, a fast computing way to approximate the derivative, which does not require any more memory storage. We approximate the derivative by finite differences as follows
\begin{equation*}
\tilde{F}'(\tilde{x}) \approx \frac{2}{b-a} \frac{I_N F(x+h)-I_N F(x)}{h},
\end{equation*}
where $0<h<<1$.

As it has been seen, all the algorithms developed in Subsections \ref{Ch1COTIP} and \ref{Ch1TEDIP} are general enough. The only thing that we need to know is how many variables the interpolated function has and everything is straightforward.

\section{Reduced Basis Approach}\label{RBFAOVRBA}

The objective of this Section is to build a new polynomial which gives comparable accuracy as the interpolant built in the previous Section, but which has less memory requirements.

Suppose we are given a high degree polynomial $P^{\textbf{N}}$. The objective is to construct from it a smaller polynomial, in memory terms but not in degree, which globally values as well as the original one.

The method we are going to develop could be exported to other kinds of polynomials, but since our evaluation algorithms are designed for the Chebyshev ones, the construction is focused to take advantage of their properties. It is also general enough to be applied to any n-variables  polynomial.

\subsection{Hierarchical orthonormalization}\label{Ch1Subhierarproc}

Suppose we have a polynomial
\begin{equation}
P^{\boldsymbol{N}} \left(x_1, \ ..., \ x_n\right), \quad  \boldsymbol{N}=[N_1,N_2,..., N_n]\in\mathbb{N}^n,
\end{equation}
where $\boldsymbol{N}$ denotes the degree in each one of the variables.

Our objective is, given a set of points $\Phi=\{\phi_i\}^{m_{\Phi}}_{i=1}$ and $\epsilon >0$, to construct a polynomial $Q^{\boldsymbol{N}^{\Phi}_{\epsilon}}$ from $P^{\boldsymbol{N}}$, such that
\begin{equation}\label{Ch1forhierproc}
\frac{1}{m_{\Phi}}\sum_{i=1}^{m_{\Phi}}\left(P^{\boldsymbol{N}}(\phi_i)-Q^{\boldsymbol{N}^{\Phi}_{\epsilon}}(\phi_i)\right)^2 < \epsilon,
\end{equation}
where polynomial $Q^{\boldsymbol{N}^{\Phi}_{\epsilon}}$ has the smallest size (in memory terms) compatible with (\ref{Ch1forhierproc}).

Although another set of points could be chosen, since we are continuing the work of the previous Section, i.e. $P^{\boldsymbol{N}}=I_{\boldsymbol{N}}F$, the interpolation polynomial of a certain function $\tilde{F}(\tilde{\boldsymbol{x}})$, the natural set of points $\Phi$ will be the set of points used in the construction of the interpolation polynomial, i.e., the Chebyshev nodes $\Phi=\{\boldsymbol{\alpha}_{\boldsymbol{l}}\}_{\boldsymbol{l}\in L^{\boldsymbol{N}}}$.

Our approach is to use a basis of Orthonormal functions that are hierarchically chosen. All polynomials that appear in the procedure we are going to construct must be written in function of Chebyshev polynomials. Hence, it is natural to employ the weighted norm associated with them and to exploit all the related properties which simplify the calculus.

\begin{definition}{\label{Ch1defprodesccheby}}
Given two functions $f(x_1,...,x_n)$ and $g(x_1,...,x_n)$, where $(x_1,...,x_n)\in [-1, \ 1]^{n}$, we define the weighted scalar product $<f,g>_{\mathrm{L_{\omega}}}$ as:
\begin{equation*}
<f,g>_{\mathrm{L_{\omega}}}=\int_{-1}^1 ... \int_{-1}^1 \frac{f(x_1,...,x_n)g(x_1,...,x_n)}{{\sqrt{1-x_1^2}}{...}{\sqrt{1-x_n^2}}} dx_1 ... dx_n.
\end{equation*}
We denote by $||\cdotp||_{\mathrm{L_{\omega}}}$ the norm induced by this scalar product.
\end{definition}

The following result is well known \cite{Rivlin}.

\begin{lemma}{\label{Ch1lemprodesccheby}}
The Chebyshev polynomials, $T_i(x)$, are orthogonal with respect to the scalar product $<f,g>_{\mathrm{L_{\omega}}}$. Furthermore, if $H(x)$ is a polynomial of degree less or equal to $2n+1$ then
\begin{equation*}
\int^{1}_{-1} \frac{H(x)}{{\sqrt{1-x^2}}} dx=\sideset{_{ }^{ }}{_{ }^{''}}\sum_{j=0}^n \frac{\pi}{n}H(x_j)=\frac{\pi}{2n}H(x_0)+\sum_{j=1}^{n-1} \frac{\pi}{n}H(x_j) +\frac{\pi}{2n}H(x_n),
\end{equation*}
where $\{x_j\}_{j=0}^{n}$ are the Chebyshev nodes in [-1,1]. (Gauss-Lobato-Chebyshev cuadrature)
\end{lemma}

\begin{definition} Let $f(x_j,x_{j+1},...,x_n)$ and $g(x_j,x_{j+1},...,x_n)$ be two functions such that $(x_j,x_{j+1}...,x_n)\in [-1, \ 1]^{n-j+1}$. We define the function
\begin{equation*}
<f, g>_{\mathrm{L^{j+1,n}_{\omega}}}(x_j)= \int_{-1}^1 ... \int_{-1}^1 \frac{f(x_j,...,x_n)g(x_j,...,x_n)}{{\sqrt{1-x_{j+1}^2}}{...}{\sqrt{1-x_n^2}}} dx_{j+1} ... dx_n.
\end{equation*}
For simplicity in the notation, we denote $<f, g>_{\mathrm{L^{j+1,n}_{\omega}}}(x_j)=<f, g>_{\mathrm{L^{j+1,n}_{\omega}}}$.
\end{definition}

The algorithm of the hierarchical Gram-Schmidt procedure has $n-1$ steps if the polynomial $P^{\boldsymbol{N}}(x_1, \ ..., \ x_n)$ has $n$ variables.

\

\noindent \textbf{Hierarchical orthonormalization procedure:}

Let us consider $P^{\boldsymbol{N}}(x_1, \ ..., \ x_n)$. For simplicity assume that $x_j\in[-1, \ 1]$. Let us consider also a set of points $\Phi=\{\phi_i\}^{m_{\Phi}}_{i=1}, \ \phi_i\in[-1,1]^{n}$.

\

\noindent \textbf{Step 1:}

Let $\{\alpha^i_1\}^{N_1}_{i=0}$ denote the $N_1+1$ Chebyshev nodes in $[-1, \ 1]$ and define $P_i(x_2,...,x_n)=P^{\boldsymbol{N}}(\alpha^i_1,x_2,...,x_n)$. It is easy to check that we can rewrite $P^{\boldsymbol{N}}$ as:
\begin{equation*}
P^{\boldsymbol{N}}(x_1, \ ..., \ x_n)=\sum^{N_1}_{i=0}a_i(x_1)P_i(x_2,...,x_n)=R_1(x_1,...,x_n),
\end{equation*}
where $a_i(x_1)$ is a $N_1$-degree polynomial such that for $m=0,1,...,N_1$ it holds that:
\begin{equation*}
\left\{
\begin{aligned}
& a_i(\alpha^i_1)=1,  \\
& a_i(\alpha^m_1)=0, \quad i\neq m.
\end{aligned}
\right.
\end{equation*}

For $i_1\in\{0,...,N_1\}$, we set $\tilde{q}^1_{i_1}=P_{i_1}$ and $q^1_{i_1}=\frac{\tilde{q}^1_{i_1}}{\left\|\tilde{q}^1_{i_1}\right\|_{L_{\omega}}}$ and compute:
\begin{equation*}
j_1=\underset{i_1}{\text{argmin}}\left\|R_1-<R_1,q^1_{i_1}>_{L^{2,n}_{\omega}}q^1_{i_1}\right\|_{L_{\omega}}.
\end{equation*}

We define $q_{j_1}=q^1_{j_1}$ and $R_2=R_1-<R_1,q_{j_1}>_{L^{2,n}_{\omega}}q_{j_1}$.

For $i_2\in\{0,...,N_1\}-j_1$, we set $\tilde{q}^2_{i_2}=P_{i_2}-<P_{i_2},q_{j_1}>_{L^{2,n}_{\omega}}q_{j_1}$ and $q^2_{i_2}=\frac{\tilde{q}^2_{i_1}}{\left\|\tilde{q}^2_{i_1}\right\|_{L_{\omega}}}$ and compute
\begin{equation*}
j_2=\underset{i_2}{\text{argmin}}\left\|R_2-<R_2,q^2_{i_2}>_{L^{2,n}_{\omega}}q^2_{i_2}\right\|_{L_{\omega}}.
\end{equation*}

We define $q_{j_2}=q^2_{j_2}$ and $R_3=R_2-<R_2,q_{j_2}>_{L^{2,n}_{\omega}}q_{j_2}$.

We proceed iteratively, so that we eventually obtain a set of orthonormal polynomials $\{q_{j_k}\}_{k=0}^{N_1}$ such that
\begin{equation*}
P^{\boldsymbol{N}}(x_1, \ ..., \ x_n)=\sum^{N_1}_{k=0}A^1_{j_k}(x_1)q_{j_k}(x_2,...,x_n),
\end{equation*}
where $A^1_{j_k}(x_1)=<P^{\boldsymbol{N}},q_{j_k}>_{L^{2,n}_{\omega}}, \ k=0,1,...,N_1$.

We approximate now
\begin{equation*}
P^{\boldsymbol{N}}(x_1, \ ..., \ x_n)\approx \sum^{M_1}_{k=0}A^1_{j_k}(x_1)q_{j_k}(x_2,...,x_n)=Q_1(x_1, \ ..., \ x_n),
\end{equation*}
where $M_1$ is the first index such that
\begin{equation}\label{Ch1sentidorecorte}
\frac{1}{m_{\Phi}}\sum_{i=1}^{m_{\Phi}}\left(P^{\boldsymbol{N}}(\phi_i)-Q_1(\phi_i) \right)^2 < \epsilon.
\end{equation}

Let us observe that the first polynomials $q_{j_k}$ were those that, in the sense of (\ref{Ch1sentidorecorte}), had more ``information'' about $P^{\boldsymbol{N}}$. Indeed, usually with very few terms (depending on the variable), a good approximation to the original polynomial can be achieved.

Furthermore, the amount of storage required is considerably reduced.

\

\noindent \textbf{Step 2:}

Each of the $q_{j_k}(x_2,...,x_n)$ is a $n-1$ variable polynomial, and we can proceed the same way as we did in \textbf{Step 1}.

\textit{For each of the $j_k$}, let $\{\alpha^i_2\}^{N_2}_{i=0}$ be the $N_2+1$ Chebyshev nodes in $[-1,1]$ and
\begin{equation*}
q_{j_k}(x_2,...,x_n)=\sum^{N_2}_{i=0}a^{j_k}_{i}(x_2)P^{j_k}_{i}(x_3,...,x_n)=R^{j_k}_{1}(x_2,...,x_n),
\end{equation*}
where $P^{j_k}_{i}(x_3,...,x_n)=q_{j_k}(\alpha^i_2,x_3,...,x_n)$.

For $i_1\in\{0,...,N_2\}$, we set $\tilde{q}^{j_k,1}_{i_1}=P^{j_k}_{i}$ and $q^{j_k,1}_{i_1}=\frac{\tilde{q}^{j_k,1}_{i_1}}{\left\|\tilde{q}^{j_k,1}_{i_1}\right\|_{L_{\omega}}}$ and compute:
\begin{equation*}
l_1=\underset{i_1}{\text{argmin}}\left\|R^{j_k}_{1}-<R^{j_k}_{1},q^{j_k,1}_{i_1}>_{L^{3,n}_{\omega}}q^{j_k,1}_{i_1}\right\|_{L_{\omega}}.
\end{equation*}

We define $q_{j_k,l_1}=q^{j_k,1}_{l_1}$ and $R^{j_k}_{2}=R^{j_k}_{1}-<R^{j_k}_{1},q_{j_k,l_1}>_{L^{3,n}_{\omega}}q_{j_k,l_1}$.

Now, for $i_2\in\{0,...,N_2\}-l_1$, we set $\tilde{q}^{j_k,2}_{i_2}=P^{j_k}_{i_2}-<P^{j_k}_{i_2},q_{j_k,l_1}>_{L^{3,n}_{\omega}}q_{j_k,l_1}$ and $q^{j_k,2}_{i_2}=\frac{\tilde{q}^{j_k,2}_{i_2}}{\left\|\tilde{q}^{j_k,2}_{i_2}\right\|_{L_{\omega}}}$. Again we compute
\begin{equation*}
l_2=\underset{i_2}{\text{argmin}}\left\|R^{j_k}_{2}-<R^{j_k}_{2},q^{j_k,2}_{i_2}>_{L^{3,n}_{\omega}}q^{j_k,2}_{i_2}\right\|_{L_{\omega}}.
\end{equation*}

We define $q_{j_k,l_2}=q^{j_k,2}_{l_2}$ and $R^{j_k}_3=R^{j_k}_2-<R^{j_k}_2,q_{j_k,l_2}>_{L^{3,n}_{\omega}}q_{j_k,l_2}$.

We proceed iteratively, at the end we will obtain
\begin{equation*}
Q_1=\sum^{M_1}_{k=0}A^1_{j_k}(x_1)\left(\sum^{N_2}_{m=0}A^2_{j_k,l_m}(x_2)q_{j_k,l_m}(x_3,...,x_n)\right),
\end{equation*}
where $A^2_{j_k,l_m}(x_2)=<q_{j_k}(x_2,...,x_n),q_{j_k,l_m}(x_3,...,x_n)>_{L^{3,n}_{\omega}}, \ m=0,1,...,N_2$.

We will approximate $P^{\boldsymbol{N}}(x_1, \ ..., \ x_n)$ by
\begin{equation*}
P^{\boldsymbol{N}}\approx \sum^{M_1}_{k=0}A^1_{j_k}(x_1)\left(\sum^{M_2}_{m=0}A^2_{j_k,l_m}(x_2)q_{j_k,l_m}(x_3,...,x_n)\right)=Q_2(x_1,...,x_n),
\end{equation*}
where $M_2$ is the first index such that
\begin{equation}
\frac{1}{m_{\Phi}}\sum_{i=0}^{m_{\Phi}}\left(P^{\boldsymbol{N}}(\phi_i)-Q_2(\phi_i)\right)^2 < \epsilon.
\end{equation}

We proceed iteratively until completing \textbf{Step n-1} where we stop. We will have arrived to a new polynomial that can be written $P^{\boldsymbol{N}}\approx Q^{\boldsymbol{N}^{\Phi}_{\epsilon}}=$
\begin{equation*}
\sum^{M_1,M_2,...,M_{n-1}}_{i_1,i_2,...,i_{n-1}}A^1_{i_1}(x_1)A^2_{i_1,i_2}(x_2)...A^{n-1}_{i_1,i_2,...,i_{n-1}}(x_{n-1})q^{n-1}_{i_1,i_2,...,i_{n-1}}(x_n).
\end{equation*}

Note that ${\boldsymbol{N}^{\Phi}_{\epsilon}}=\{M_1,...,M_{n-1},N_n\}$ is the number of function basis used in each variable. Note also that, in general, the degree of $Q^{\boldsymbol{N}^{\Phi}_{\epsilon}}$ is still $N_1\times N_2 \times...\times N_n$.

We remark that the last value of ${\boldsymbol{N}^{\Phi}_{\epsilon}}$ is $N_n$ because the last variable remains untouched. An improved result (in memory terms) can be obtained if the variables of polynomial $P^{\boldsymbol{N}}$ are reordered before the Reduced Basis procedure and $N_n$ is the smallest among the $N_i$.

\begin{figure}[h]
\centering
\includegraphics[width=13.5 cm,height=8.5cm]{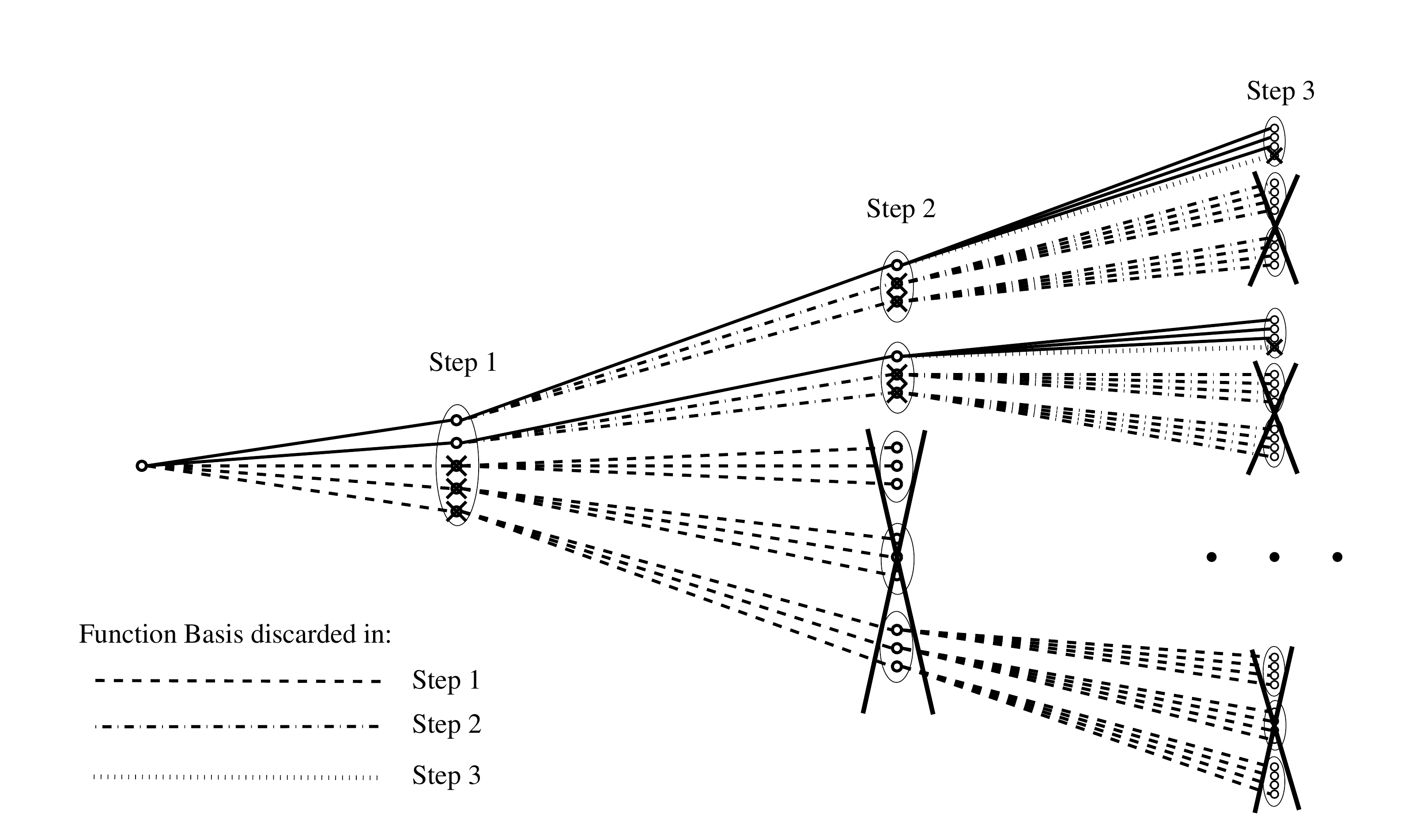}
\caption[Truncated Orthonormal decomposition]{Truncated Orthonormal decomposition. The Figure shows the function basis discarded (equivalent to memory savings) in each step (Step 1-dashed lines, Step 2-dash-dot lines and Step 3-dot lines) of the Hierarchical orthonormalization procedure. \label{Ch1truncacion3etapa}}
\end{figure}

Visually, we can check the big memory saving. Figure \ref{Ch1truncacion3etapa} shows an example of the first three steps of the algorithm if $N_1=5, \ N_2= 3, \ N_3=4, ...$.

The tree represents the orthonormal decomposition in each of the variables and the function basis that are discarded in each step: 3 function basis (Step 1-dashed lines), 2 function basis (Step 2-dash-dot lines) and 1 function basis (Step 3-dot lines). The proportion of the tree that has been discarded represents approximately the memory savings with respect to the original polynomial $P_N$ obtained by our method.

The last step is to adequate the algorithms for tensorial evaluation to $Q^{\boldsymbol{N}^{\Phi}_{\epsilon}}$.

\subsection{Tensorial evaluation for Reduced Basis}

Polynomial $Q^{\boldsymbol{N}^{\Phi}_{\epsilon}}$ is rewritten for tensorial evaluation as
\begin{equation*}
\begin{aligned}
Q^{\boldsymbol{N}^{\Phi}_{\epsilon}} &=\sum^{M_1,M_2,...,M_{n-1}}_{i_1,i_2,...,i_{n-1}}A^1_{i_1}(x_1)A^2_{i_1,i_2}(x_2)...A^{n-1}_{i_1,i_2,...,i_{n-1}}(x_{n-1})q^{n-1}_{i_1,i_2,...,i_{n-1}}(x_n) \\
&=\sum^{M_1}_{i_1=0}A^1_{i_1}(x_1)\sum^{M_2}_{i_2=0}A^2_{i_{1,2}}(x_{2})...\sum^{M_{n-2}}_{i_{n-2}=0}A^{n-2}_{i_{1,n-2}}(x_{n-2}) \\
& \ \ \sum^{M_{n-1}}_{i_{n-1}=0}A^{n-1}_{i_{1,n-1}}(x_{n-1})q^{n-1}_{i_{1,n-1}}(x_n).
\end{aligned}
\end{equation*}
where we denote $i_{1,j}=(i_1,i_2,...,i_j)$.

Each polynomial is written in function of the Chebyshev polynomials:
\begin{equation}\label{Ch1polinqstore}
\begin{aligned}
A^1_{i_1}(x_1) &=\sum_{j=0}^{N_1}a_{j,i_1}^1T(x_1), && A^2_{i_{1,2}}(x_2) =\sum_{j=0}^{N_2}a_{j,i_{1,2}}^2T(x_2), \quad ... \\
A^{n-1}_{i_{1,n-1}}(x_{n-1})&=\sum_{j=0}^{N_{n-1}}a_{j,i_{1,n-1}}^{n-1}T(x_{n-1}), && q^{n-1}_{i_{1,n-1}}(x_n)=\sum_{j=0}^{N_{n}}a_{j,i_{1,n-1}}^{n}T(x_{n}), \\
\end{aligned}
\end{equation}
and $Q^{\boldsymbol{N}^{\Phi}_{\epsilon}}$ is kept in memory storing the coefficient of the previous polynomials in $n$ different multidimensional arrays.

Suppose that we want to evaluate the polynomial in the finite set of points $\boldsymbol{\Theta}$ (see (\ref{defconjtheta})).

\

We employ the notation $\eta_j=\{x^k_j\}_{k=1}^{q_j}$, whose values are obtained from set $\boldsymbol{\Theta}$ after the change of variable given by (\ref{Ch1camcoordmult}).

The evaluation of polynomials $A^1_{i_1}(\eta_1)$, ..., $A^{n-1}_{i_{1,n-1}}(\eta_{n-1})$, $q^{n-1}_{i_{1,n-1}}(\eta_n)$ when they are given by (\ref{Ch1polinqstore}) can be done efficiently using the first algorithm from Subsection \ref{Ch1TEDIP}.

Thus, suppose that we have evaluated them and stored the results in arrays:
\begin{equation*}
\begin{aligned}
A^1_{i_1}(\eta_1) &=(A^1)_{(M_1+1)\times q_1},  \\
A^2_{i_{1,2}}(\eta_2) &=(A^2)_{(M_1+1)\times (M_2+1) \times q_2}, \\
 & ...  \\
A^{n-1}_{i_{1,n-1}}(\eta_{n-1})&=(A^{n-1})_{(M_1+1)\times (M_2+1) \times ... \times (M_{n-1}+1) \times q_{n-1}},   \\
q^{n-1}_{i_{1,n-1}}(\eta_n)&=(A^{n})_{(M_1+1)\times  (M_2+1) \times ... \times (M_{n-1}+1) \times q_{n}}.  \\
\end{aligned}
\end{equation*}

\begin{definition}\label{DEfCh1esptensarrop}
Let $A$, $B$ be two arrays such that $A=(A)_{m_1\times m_2 \times ... \times m_k \times a }$ and $B=(B)_{m_1\times m_2 \times ... \times m_k \times b_1 \times ... \times b_s}$.

We define the special tensorial array operation $C=A\tilde{\otimes} B$ as:
\begin{equation}\label{Ch1esptensarrop}
\begin{aligned}
&C(j_1,...,j_{k-1},:,j_{k+1},...,j_{k+s}) \\
& \ \ \ =A(j_1,...,j_{k-1},:,:)'  B(j_1,...,j_{k-1},:,j_{k+1},...,j_{k+s}),
\end{aligned}
\end{equation}
where $\cdotp$ denotes the usual product of matrix times a vector.
\end{definition}

From Definition \ref{DEfCh1esptensarrop}, observe that in (\ref{Ch1esptensarrop}) we are using the usual matrix times vector multiplication.
It is straightforward that $\dim(C)=m_{1}\times m_{2}\times ... \times m_{k-1} \times a \times b_1 \times ... \times b_{s}$.

\textit{multiprod} command is employed again to implement the special tensorial array operation.

The tensorial evaluation for the reduced basis polynomial can be written as:
\begin{equation*}
Q^{\boldsymbol{N}^{\Phi}_{\epsilon}}=A^1_{i_1}(\eta_1)\tilde{\otimes}\left(...\tilde{\otimes}\left(A^{n-2}_{i_{1,n-2}}(\eta_{n-2})\tilde{\otimes}\left(A^{n-1}_{i_{1,n-1}}(\eta_{n-1})\tilde{\otimes}q^{n-1}_{i_{1,n-1}}(\eta_n)\right)\right)...\right),
\end{equation*}
where again the order fixed by the parenthesis must be strictly followed in order to be consistent with the dimensions of the arrays.

The result will be a $q_1\times q_2 \times ... \times q_n$-dimensional array which contains the evaluation of the polynomial with all the possible combinations of the given values to each of the variables.

\subsection{Comments about the Reduced Basis method.}

Although in the numerical experiments we will see that the results are quite good in the sense of reduction of memory requirements and computing time, we must point that the procedure presented could be improved.

We remark that $Q^{\boldsymbol{N}^{\Phi}_{\epsilon}}$ is not optimal in various senses. First of all, the hierarchical criteria to select the function basis is not necessarily optimal. For example, there might exist a combination of several function basis that give a less overall error than the ones chosen by our criterium. The selection criterium that we employ is fast because, when we have to order hierarchically the function basis in each step, we only need to reevaluate the function basis that have not already been ordered.

Another factor that could be improved is the criterium for truncation. We can orthonormally decompose the whole polynomial hierarchically
\begin{equation*}
P^{\boldsymbol{N}}=\sum^{N_1,N_2,...,N_{n-1}}_{i_1,i_2,...,i_{n-1}}A^1_{i_1}(x_1)A^2_{i_1,i_2}(x_2)...A^{n-1}_{i_1,i_2,...,i_{n-1}}(x_{n-1})q^{n-1}_{i_1,i_2,...,i_{n-1}}(x_n),
\end{equation*}
and notice that we can independently truncate one branch of the tree or another. For example:
\begin{equation*}
P^{\boldsymbol{N}}\approx\sum^{M_1,N_2,N_3...,N_{n-1}}_{i_1,i_2,...,i_{n-1}}A^1_{i_1}(x_1)A^2_{i_1,i_2}(x_2)...A^{n-1}_{i_1,i_2,...,i_{n-1}}(x_{n-1})q^{n-1}_{i_1,i_2,...,i_{n-1}}(x_n),
\end{equation*}
or
\begin{equation*}
 P^{\boldsymbol{N}}\approx\sum^{N_1,M_2,N_3,...,N_{n-1}}_{i_1,i_2,...,i_{n-1}}A^1_{i_1}(x_1)A^2_{i_1,i_2}(x_2)...A^{n-1}_{i_1,i_2,...,i_{n-1}}(x_{n-1})q^{n-1}_{i_1,i_2,...,i_{n-1}}(x_n).
\end{equation*}

Our procedure does not maximize memory savings over the whole polynomial. A method which maximizes the memory savings versus the deterioration of the error when we truncate function basis of one or other variable could be designed.

\section{Numerical Results}\label{RBFAOVNR}

We are now going to apply the techniques developed in Sections \ref{RBFAOVPi} and \ref{RBFAOVRBA} to a multidimensional model employed in option pricing.

The outline of the Section is as follows. First, we will make a brief introduction about financial options and pricing models. Afterwards, we will build an interpolation polynomial of a particular pricing function and apply the Reduced Basis approach. Performance analysis when we employ both numerical approximations will be performed.

\subsection{Option Pricing and GARCH models}\label{RBFAOVGMNm}

An European Call Option is a financial instrument that gives the buyer the right, but not the obligation, to buy a stock or asset, at a fixed future date (maturity), and at a fixed price (strike or exercise price). The seller will have the obligation, if the buyer exercises his/her right, to sell the stock at the exercise price.

The stock is usually modelled as an stochastic process and empirical analysis show that the volatility of the process does not remain constant in time. ARCH models (AutoRegressive Conditional Heterodastic) introduced by Engle in \cite{Engle1} are a kind of stochastic processes in which recent past gives information about future variance. Several ARCH models have been proposed along the years, trying to capture some of the empirically observed stock properties. The objective of this work is not to give a deep review of the ARCH literature, and we refer to \cite{Bollerslev1}, \cite{Bollerslev2}, \cite{Christoffersen}, \cite{Higgins} and references therein. Nevertheless, we point that ARCH models are broadly used in option pricing. \cite{Breton}, \cite{Duan3}, \cite{Harrel}, \cite{Ritchen} and \cite{Stentoft} are just a few examples where option prices are obtained through ARCH models.

In the present work, we are going to apply the techniques that we have developed to price options with the NGARCH(1,1) model. For pricing options, the dynamics of the stock in the risk-free measure $Q$ of the NGARCH(1,1) model \cite{Duan1}, \cite{Kallsen} are,
\begin{equation}\label{Ch1ecuGrachQ}
   \left\{
    \begin{aligned}
         \ln\left(\frac{S_{t}}{S_{t-1}}\right)\equiv R_t & =r-\frac{1}{2}\sigma^2_t+\sqrt{\sigma^2_t}z^{Q}_t, \\
         \sigma^2_t & =\beta_0+\beta_1\sigma^2_{t-1}+\beta_2\sigma^2_{t-1}(z^{Q}_{t-1}-(\theta+\lambda))^2,
    \end{aligned}
   \right.
\end{equation}
where $S_t$ denotes the stock price, $\sigma^2_t$ is the variance of the stock, $r$ is the risk-free rate, $\beta_0, \ \beta_1, \ \beta_2, \ \lambda, \ \theta$ are the GARCH model parameters and $z^{Q}_t$ is a normally distributed random variable with mean 0 and variance 1.

The dynamics in the Risk-free measure allow us to compute the European option price as:
\begin{equation}\label{Ch1ecuGrachQ2}
   C(S)=e^{-r({t_M}-t)}\mathbb{E}^Q\left[\max{\{S_{t_M}-K,0\}}|S_t=S\right].
\end{equation}
which, taking into account the model parameters, is an 8-variable function.

For this model, there is no known closed form solution and several numerical methods can be employed, being Monte-Carlo based methods (\cite{Duan2}, \cite{Stentoft}), Lattice methods (\cite{Lyuu}, \cite{Ritchen}), Finite Elements (\cite{Achdou}) or Spectral methods (\cite{Breton}) some of them.

The principal drawback of GARCH models is their computational cost. It will depend on the numerical method employed, but all the ones mentioned (Monte-Carlo, Lattice, Spectral...) require several seconds to compute option prices and several minutes to estimate parameter values. This can result in an unpractical procedure, since option prices change almost continuously.

In this work, the numerical method employed for computing the option prices in the Chebyshev nodes will be the spectral method developed in \cite{Breton} and will be referred as B-F method. This numerical method gives enough precision with few grid points, and the employment of FFT techniques makes it a low-time consuming method.

Fixed an enough precision for the B-F method, we assume for the rest of the work that the option price obtained with this method is the reference option price. The construction of the interpolating polynomial and the error analysis will be carried referencing to the values obtained with it.

\subsection{Numerical analysis of Interpolation}\label{RBFAOVPiIE}

First, we fix the intervals in which the interpolant of the option price will be constructed. The NGARCH(1,1) model is linear in the relation $\frac{S}{K}$, so strike is fixed at $K=1$. The rest of variables are defined as follows:
\begin{equation*}
\begin{aligned}
& t_M \in [0, \ 365], \ \ \ \ \ \ \  &&\beta_0 \in [0, \ 2\cdotp10^{-6}],  \\
& h_0 \in [0\ldotp25\cdotp10^{-4}, \ 2\ldotp25\cdotp10^{-4}], \ &&\beta_1 \in [0\ldotp60, \ 0\ldotp95], \\
& S_0 \in [0\ldotp75, \ 1\ldotp20], \ \ &&\beta_2 \in [0\ldotp02, \ 0\ldotp25], \\
& r \in [0\ldotp02, \ 0\ldotp085], \ \ &&(\lambda+\theta) \in [0\ldotp20, \ 2].
\end{aligned}
\end{equation*}
where these intervals are chosen because they cover usual parameter values of the model observed in the literature (see, for example, \cite{Christoffersen}, \cite{Duan3} or \cite{Stentoft}).

Although different number of nodes can be considered for each variable, for simplicity, consider $\boldsymbol{N}=(N,N,...,N)$.

We are going to carry out a standard error analysis doubling the number of nodes $N=3,6,12$. We remark that the number of interpolation points of variable $x_{j}$ is $N_{j}+1$ and the storage cost of the polynomials is $8\prod_{j=1}^n N_{j}+1$.

Once fixed $\boldsymbol{N}$, we compute the Chebyshev nodes $\{\tilde{\boldsymbol{\alpha}}^{\boldsymbol{l}}\}_{l\in L^{\boldsymbol{N}}}$  with formula (\ref{Ch1forChbnodes}). We compute the function values $\{\tilde{F}\left(\tilde{\boldsymbol{\alpha}}^{\boldsymbol{l}}\right)\}_{l\in L^{\boldsymbol{N}}}$ with B-F method and construct $I_{\boldsymbol{N}} F$ with the algorithms developed in Subsection \ref{Ch1COTIP}.

Independently, we have to build a control sample which allows us to measure how well the interpolation polynomial prices options in the domain $\tilde{\Omega}$. We have chosen a set uniformly defined over $\tilde{\Omega}$.

\begin{definition}
For each variable $\tilde{x}_j\in [\tilde{x}^{\min}_j, \ \tilde{x}^{\max}_j]$ and for a given $m\in\mathbb{N}$ we define
\begin{equation*}
\Delta^m_{\tilde{x}_j}=\frac{\tilde{x}^{\max}_j-\tilde{x}^{\min}_j}{m},
\end{equation*}
and the set of points
\begin{equation*}
\Theta^m_{\tilde{x}_j}=\left\{\tilde{x}^{\min}_j+\Delta_{\tilde{x}_j}  i \right\}_{i=1}^{m-1}, \quad j=1,2...,8.
\end{equation*}
\end{definition}

This set of equally spaced points will be used to build a control sample. Points that correspond to $i=\{0,m\}$ are not included because they always correspond to Chebyshev nodes. $\text{Sample }\boldsymbol{\Theta}_m$ will denote the set of option prices for all the possible combinations of values in sets $\Theta^m_{\tilde{x}_j}$, i.e.
\begin{equation*}
\text{Sample }\boldsymbol{\Theta}_m=\{\tilde{F}\left(\Theta^m_{\tilde{x}_1},\Theta^m_{\tilde{x}_2},...,\Theta^m_{\tilde{x}_8}\right)\}, \quad |\text{Sample }\boldsymbol{\Theta}_m|=(m-1)^8,
\end{equation*}
computed with B-F method.

We compute $\text{Sample }\boldsymbol{\Theta}_m$ and numerate its elements. Let $C^{B-F}_j$ be the $j$ contract price of $\text{Sample }\boldsymbol{\Theta}_m$ and $C^{I_{\boldsymbol{N}}F}_j$ be the $j$ contract price evaluated with polynomial $I_{\boldsymbol{N}}F$.

We define
\begin{equation*}
\text{MSE}_{\text{Sample }\boldsymbol{\Theta}_m}(I_{\boldsymbol{N}}F)=\frac{1}{(m-1)^8}\sum_{j=1}^{(m-1)^8}\left(C^{B-F}_{j}-C^{I_{\boldsymbol{N}}F}_j\right)^2.
\end{equation*}

For the numerical examples, we have built Sample $\boldsymbol{\Theta}_7$. The number of elements of this sample is $1679616\approx 1\ldotp68 \cdotp 10^6$.

Table \ref{Ch3tableI3I6I12} shows, for $\boldsymbol{N}=\boldsymbol{3},\boldsymbol{6},\boldsymbol{12}$,  the memory storage (bytes) requirements of $I_{\boldsymbol{N}}F$, the computing time (seconds) of computing Sample $\boldsymbol{\Theta}_7$ with $I_{\boldsymbol{N}}F$ and the $\text{MSE}_{\text{Sample }\boldsymbol{\Theta}_7}(I_{\boldsymbol{N}}F)$.

\begin{table}[h]
\centering
    \begin{tabular}{|c|c|c|c|}
    \hline
                             & Storage        & Computing time (seconds)  & $\text{MSE}_{\text{Sample }\boldsymbol{\Theta}_7}(I_{\boldsymbol{N}}F)$   \\ \hline
    $I_{\boldsymbol{3}}F$    & $5\ldotp24 \cdotp10^{5}$      & $0\ldotp11$                     & $0\ldotp6918 \cdotp 10^{-4}$    \\
    $I_{\boldsymbol{6}}F$    & $4\ldotp61 \cdotp10^{7}$      & $0\ldotp46$                     & $0\ldotp1229 \cdotp 10^{-4}$    \\
    $I_{\boldsymbol{12}}F$   & $6\ldotp52 \cdotp10^{9}$      & $91$                            & $0\ldotp0120 \cdotp 10^{-4}$    \\ \hline
    \end{tabular}
    \caption{ \label{Ch3tableI3I6I12} Storage cost of $I_{\boldsymbol{N}}F$, computing time for evaluating $\text{Sample }\boldsymbol{\Theta}_7$ with $I_{\boldsymbol{N}}F$ and the Mean Square Error committed by the interpolation polynomial $I_{\boldsymbol{N}}F$ when evaluating the contract prices of $\text{Sample }\boldsymbol{\Theta}_7$.}
\end{table}

In Table \ref{Ch3tableI3I6I12} we can also check that the computing time of  $I_{\boldsymbol{3}}F$ and $I_{\boldsymbol{6}}F$ is fairly good, but it blows to 91 seconds in the case of $I_{\boldsymbol{12}}F$. Although this time might not seem too high for computing $\approx 1\ldotp68 \cdotp 10^6$ contracts, it is unacceptable for practical applications. In the markets, the stock price might have changed a few times before we have finished the computation, making the results worthless.

The reason why the computing time has increased so much is due to the ``Curse of dimensionality''. We remark that $I_{\boldsymbol{12}}F$ is above the operational limit of the Matlab/computer employed in the analysis to be stored in just one single array. Although the storage problem can be handled, splitting the polynomial in several parts and loading/discarding the needed data, unfortunately, in velocity terms, this implies a large increment of computational time.

Concerning $I_{\boldsymbol{12}}F$, let $\left\{{t_M}_i\right\}_{i=0}^{12}$ denote the 13 Chebyshev nodes in interval $[0, \ 365]$. For each value of ${t_M}_i$ we build the 7-variable interpolation polynomial for the rest of the variables. This way, we have polynomial $I_{\boldsymbol{12}}F$ stored as 13 smaller polynomials which can be handled.

In our example, the 91 seconds are mostly due to several uses of the function \textit{load} when we call each of the 7-variable smaller polynomials.

We mention that if the polynomial was even bigger, the splitting procedure can be extended to other variables, so storage is not an unsolvable problem. Nevertheless, it worsens the computing time because it implies that we need to load data from the memory very frequently.

\begin{figure}[h]
\centering
\includegraphics[width=6cm,height=5cm]{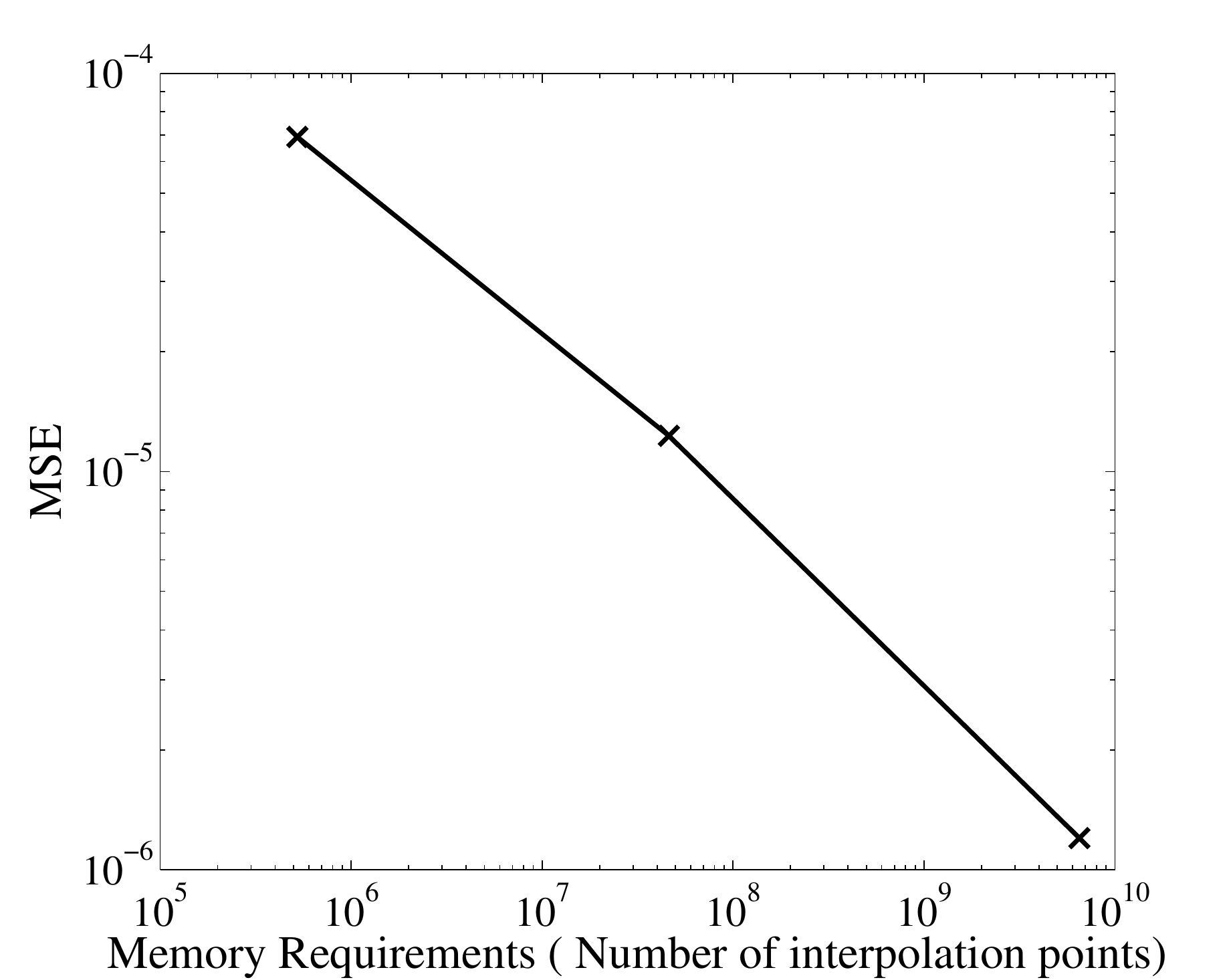}
\caption[Interpolation Error Convergence]{\label{Ch1v1converrinterp} Interpolation Error Convergence. We plot the log-log of the Memory requirements (horizontal axis) vs the Mean Square Error (vertical axis) of each polynomial $I_{\textbf{N}}F$.}
\end{figure}

Concerning the error of the interpolation polynomials, Figure \ref{Ch1v1converrinterp} shows the log-log of the Memory Storage versus the Mean Square Error.

The slope of the regression line in Figure \ref{Ch1v1converrinterp} is $-0.43$. We have a good error behaviour, achieving a precision of $0\ldotp01203 \cdotp 10^{-4}$ with $\boldsymbol{N}=\{12,12,...,12\}$. If more precision is required, we can build bigger interpolation polynomials, which can be handled thanks to the splitting technique that we previously described.

\subsection{Numerical analysis of Reduced Basis.}

We now apply the Reduced Basis procedure developed in Section \ref{RBFAOVRBA} to polynomial $I_{\boldsymbol{12}}F$ built in Subsection \ref{RBFAOVPiIE}.

The set of points employed in the Hierarchical procedure will be the set of interpolation points employed in the construction of the interpolation polynomial $I_{\boldsymbol{12}}F$, i.e. the Chebyshev nodes   $\Phi=\{\boldsymbol{\alpha}_{\boldsymbol{l}}\}_{\boldsymbol{l}\in L^{\boldsymbol{N}}}$.

We also want to know how the new polynomial computes option prices over the whole domain $\tilde{\Omega}$. In order to check this, we employ  $\text{Sample }\boldsymbol{\Theta}_7$ ($\approx 1\ldotp68 \cdotp 10^6$ contracts) defined in Subsection \ref{RBFAOVPiIE}, whose points do not correspond to those of $\Phi$ and are equally distributed over $\tilde{\Omega}$.

In Table \ref{Ch3tablestep1} we study the numerical results when just \textbf{Step 1} of the algorithm of the Hierarchical orthonormalization is applied. Once we have decomposed
\begin{equation*}
I_{\boldsymbol{12}}F \approx  \sum^{M_1}_{k=0}A^1_{j_k}(x_1)q_{j_k}(x_2,...,x_n),
\end{equation*}
for $M_1=0,1,...,12$, we can check the performance of the method both in Sample $\Phi$ and in
$\text{Sample }\boldsymbol{\Theta}_7$.

In Table \ref{Ch3tablestep1} are represented the number of function bases retained $(M_1+1)$, the total storage cost in each case and the Mean Square Error when they are employed to compute the option prices of Sample $\Phi$ and of $\text{Sample }\boldsymbol{\Theta}_7$.

\begin{table}[h]
\centering
    \begin{tabular}{|c|c|c|c|}
    \hline
     Function Basis  & Storage (bytes)  & $\text{MSE}_{\text{Sample }\Phi}$  & $\text{MSE}_{\text{Sample }\boldsymbol{\Theta}_7}$ \\ \hline
            1    &   $5\ldotp01\cdotp10^{8}$     &       $3\ldotp41 \cdotp 10^{-4}$       &     $2\ldotp545996\cdotp 10^{-4}$ \\
            2    &   $1\ldotp00\cdotp10^{9}$     &       $0\ldotp14 \cdotp 10^{-4}$       &     $0\ldotp088629\cdotp 10^{-4}$ \\
            3    &   $1\ldotp51\cdotp10^{9}$     &       $3\ldotp79 \cdotp 10^{-6}$       &     $0\ldotp024154\cdotp 10^{-4}$ \\
            4    &   $2\ldotp00\cdotp10^{9}$     &       $8\ldotp53 \cdotp 10^{-7}$       &     $0\ldotp018056\cdotp 10^{-4}$ \\
            5    &   $2\ldotp50\cdotp10^{9}$     &       $6\ldotp72 \cdotp 10^{-8}$       &     $0\ldotp012375\cdotp 10^{-4}$ \\
            6    &   $3\ldotp01\cdotp10^{9}$     &       $8\ldotp79 \cdotp 10^{-9}$       &     $0\ldotp012198\cdotp 10^{-4}$ \\
            7    &   $3\ldotp51\cdotp10^{9}$     &       $2\ldotp30 \cdotp 10^{-9}$       &     $0\ldotp012068\cdotp 10^{-4}$ \\
            8    &   $4\ldotp01\cdotp10^{9}$     &       $3\ldotp89 \cdotp 10^{-10}$      &     $0\ldotp012034\cdotp 10^{-4}$ \\
            9    &   $4\ldotp51\cdotp10^{9}$     &       $5\ldotp29 \cdotp 10^{-11}$      &     $0\ldotp012032\cdotp 10^{-4}$ \\
            10   &   $5\ldotp01\cdotp10^{9}$     &       $1\ldotp77 \cdotp 10^{-12}$      &     $0\ldotp012033\cdotp 10^{-4}$ \\
            11   &   $5\ldotp52\cdotp10^{9}$     &       $2\ldotp33 \cdotp 10^{-14}$      &     $0\ldotp012033\cdotp 10^{-4}$ \\
            12   &   $6\ldotp02\cdotp10^{9}$     &       $3\ldotp22 \cdotp 10^{-16}$      &     $0\ldotp012033\cdotp 10^{-4}$ \\
            13   &   $6\ldotp52\cdotp10^{9}$     &       $1\ldotp18 \cdotp 10^{-28}$      &     $0\ldotp012033\cdotp 10^{-4}$ \\ \hline
            $I_{\boldsymbol{12}}F$ & $6\ldotp52 \cdotp10^{9}$ & 0 & $0\ldotp0120330555\cdotp 10^{-4}$ \\ \hline
    \end{tabular}
    \caption{\label{Ch3tablestep1} Number of function bases retained after \textbf{Step 1} of the Hierarchical orthonormalization. We include the storage costs in each case and the MSE committed when evaluating Sample $\Phi$ and $\text{Sample }\boldsymbol{\Theta}_7$.}
\end{table}

The global error is represented by $\text{MSE}_{\text{Sample }\boldsymbol{\Theta}_7}$. Note that with just 5 or 6 function bases for the first variable, we obtain a polynomial which has comparable accuracy as $I_{\boldsymbol{12}}F$ but which requires half the storage cost.

We run now the Hierarchical orthonormalization algorithm completely (\textbf{Steps 1-7}). We fix three different values of the parameter $\epsilon$, where $\epsilon$ is the maximum Mean Square Error allowed when we evaluate Sample $\Phi$ with the polynomials obtained from the procedure.

In Table \ref{Ch3resulHAI12}, we include the memory requirements of each polynomial and the error committed when they are employed to compute the contracts of $\text{Sample }\boldsymbol{\Theta}_7$.

\begin{table}[h]
\centering
    \begin{tabular}{|c|c|c|c|c|}
    \hline
     & $\epsilon$ & MSE($\text{Sample}\boldsymbol{\Theta}_7$) & Storage (bytes) & Memory Savings \\ \hline
     $I_{\boldsymbol{12}}F$ &  &  $0\ldotp01203\cdotp 10^{-4}$ & $6\ldotp52 \cdotp10^{9}$  &   \\ \hline
     $Q_1$ & $4\cdotp 10^{-7}$ & $0\ldotp01194\cdotp 10^{-4}$ & $6\ldotp263 \cdotp10^{7}$ & $99\ldotp038$ \% \\
     $Q_2$ & $5\ldotp5 \cdotp 10^{-7}$ & $0\ldotp01229\cdotp 10^{-4}$ & $4\ldotp431 \cdotp10^{7}$ & $99\ldotp319$ \% \\
     $Q_3$ & $6\cdotp 10^{-7}$ & $0\ldotp01249\cdotp 10^{-4}$ & $3\ldotp325 \cdotp10^{7}$ & $99\ldotp489$ \% \\ \hline
    \end{tabular}
    \caption{\label{Ch3resulHAI12}Mean Square Error committed when evaluating $\text{Sample }\boldsymbol{\Theta}_7$ with three different polynomials constructed from $I_{\boldsymbol{12}}F$ after applying the Hierarchical orthonormalization procedure. We include the storage cost and the memory savings with respect to the storage cost of $I_{\boldsymbol{12}}F$.}
\end{table}

Table \ref{Ch3resulHAI12} shows that with the Reduced Bases approach we obtain much smaller polynomials (in memory terms) which give an overall error of the same order as $I_{\boldsymbol{12}}F$. We remark that, as expected, if $\epsilon\rightarrow 0$, the error committed when evaluating $\text{Sample }\boldsymbol{\Theta}_7$ converges to $0\ldotp01203\cdotp 10^{-4}$, the interpolation error of $I_{\boldsymbol{12}}F$.

Concerning the computing time, consider the sets of points: 1 contract (one value for each variable), 210 contracts (7 different stock prices, 6 volatilities, 5 maturities) and $\text{Sample }\boldsymbol{\Theta}_7$  (1679616 contracts).

We remark that evaluate $\text{Sample }\boldsymbol{\Theta}_7$ would be equivalent to price options in the real market for several stocks with different parameter values.

In Table \ref{Ch3Iq123}, we show the computing time of evaluating these sets of contracts with B-F method, the interpolation polynomial $I_{\boldsymbol{12}}F$ and different polynomials constructed from $I_{\boldsymbol{12}}F$ with the Reduced Bases approach.

\begin{table}[h]
\centering
    \begin{tabular}{|c|c|c|c|}
    \hline
      & 1 contract & 210 contracts & $\text{Sample }\boldsymbol{\Theta}_7$ ($\approx 1.68\cdotp10^6$) \\ \hline
     B-F method & $41$ s & $41$ s & $3\cdot 10^{5}$ s \\
     $I_{\boldsymbol{12}}F$ & $91$ s & $91$ s & $91$ s \\
     $Q_1$ & $0\ldotp301$ s & $0\ldotp303$ s & $0\ldotp768$ s \\
     $Q_2$ & $0\ldotp216$ s & $0\ldotp218$ s & $0\ldotp583$ s \\
     $Q_3$ & $0\ldotp162$ s & $0\ldotp164$ s & $0\ldotp499$ s \\ \hline
    \end{tabular}
    \caption{\label{Ch3Iq123}Computing time of evaluating different sets of contracts with B-F method, the interpolation polynomial $I_{\boldsymbol{12}}F$ and different polynomials constructed from $I_{\boldsymbol{12}}F$ with the Reduced Bases approach.}
\end{table}

In Table \ref{Ch3Iq123}, we can see that B-F method needs the same time for computing 1 or 210 contracts (because it admits tensorial evaluation for $S, \ \sigma^2_0$ and $T$). For computing  $\text{Sample }\boldsymbol{\Theta}_7$, B-F method needs to be evaluated for each different value of $\{\beta_0, \beta_1, \beta_2, r, (\lambda+\theta)\}$, i.e. 7776 different evaluations of 41 seconds each.

$I_{\boldsymbol{12}}F$ requires the same computing time in the three examples because, as it was mentioned in Subsection \ref{RBFAOVPiIE}, the polynomial is too big and it has to be stored in different parts. The computing time is due to several employments of function \textit{load}.

With the polynomials obtained from the Reduced Bases, we have retrieved the tensorial evaluation velocity achieved when we were working with the polynomials $I_{\boldsymbol{3}}F$, $I_{\boldsymbol{6}}F$ but with more precision (compare with Table \ref{Ch3tableI3I6I12}).

Concerning the computing time of the In the Sample analysis (a least square search), we point out that while with the usual methods (Monte-Carlo, Lattice, Spectral) it takes several minutes to estimate the parameters of one negotiation day $t_0$, this can be done in a few seconds with polynomials $Q_i$.

\subsection{Model Calibration.}

For finishing the numerical analysis, we check how this technique performs when we want to apply it to calibrate model parameters or predict the price of future European Option contracts.

The experiment has two parts.  In the first one, given a set of European Call contracts that are being traded, we want to calibrate the model (In). Our objective, is to find the parameter values of the GARCH model $\{\sigma^2_{t_0}, \beta_0, \beta_1, \beta_2, (\lambda + \theta)\}$ that give the minimum mean square error (MSE) between the theoretical option prices and the traded ones. Therefore, our objective is to find, at a moment $t_0$, the parameter values that minimize the function
\begin{equation}
\text{InMSE}(t_0)=\frac{1}{N_{t_0}}\sum^{N_{t_0}}_{i=1}\left(C^i_{t_0}-C_i(S^i_{t_0},{t_M}_i,K_i)\right)^2.
\end{equation}
where $N_{t_0}$ denotes the amount of contracts negotiated at $t_{0}$ and for each contract $i$, $C^i_{t_0}$ is the market price and $C_i(S^i_{t_0},{t_M}_i,K_i)$ is the model's price. The $r_{t_0}$ can be taken, for example, as the constant risk-free rate corresponding to the US bond negotiated at $t_0$.

\

Once we have calibrated the model, we can employ the parameters obtained to predict future option prices (Out). In the market, stock prices change almost constantly. Furthermore, new contracts with different maturities and/or strikes, which where not traded previously, can be negotiated. Assuming that the rest of the model parameters have not changed, we will study how well the polynomial approximation predicts the contract prices for the new stock prices, strikes or maturities and we compare the results with the prices that were given by the market to those contracts.

\

The real option prices traded in the market are not driven by a discrete NGARCH model. There are many factors which affect the option prices and NGARCH model is just an approximation to them. In order to generate the sets (estimation and prediction) of artificial Option prices which play the role of ``market'' prices, we employ the continuous Stochastic Volatilty model developed by Heston (see \cite{HestonSV}) given by
\begin{equation}
   \left\{
    \begin{aligned}
         dS(t) &=r S(t)dt+\sqrt{v(t)}S(t)dz_1(t), \\
         dv(t) &=\kappa^{*}[\theta^{*}-v(t)]dt+\sigma^{*}\sqrt{v(t)}dz_2(t), \\
    \end{aligned}
   \right.
\end{equation}
in the risk-neutral measure and where parameter $\rho$ denotes the instant correlation between processes $z_1$ and $z_2$ (see \cite{HestonSV}).

This way, we are inducing a noise or error, since the discrete NGARCH model does not mimic completely the continuous SV model, neither in the estimation nor in the prediction. The sets of contracts that we employ as market contracts are computed with SV model.

We remark that our objective is not to study how well does NGARCH approximates SV model. Our objective is to compare the results in the estimation and prediction of the NGARCH model (B-F method) with the results of the interpolation polynomial $I_{\boldsymbol{12}}F$ and the polynomials obtained in the Reduced Basis approximation $Q_1$, $Q_2$ and $Q_3$.

\

We fix the risk free rate $r=0.05$. The risk-free rate can be considered as an observable data, for example, it can be obtained as the constant interest rate of the US-Bond Treasury Bond. We also fix values for the set $\Omega^{*}=\{\kappa^{*},\theta^{*},\sigma^{*},v(0),\rho\}$ which corresponds to the parameter values of SV model.

We compute the ``market'' option prices with SV model for a strike $K=1$, stock prices $S=0.8,0.82,0.84,...,1.18$ and maturities $T=10,40,70,...,340$ days. The market usually trades contracts for different strikes, but we recall that the NGARCH model was linear in the relation $S/K$ so it is equivalent to fix the strike and compute option prices for different stock prices.

Table \ref{Artconsist} represents the estimation of the parameter values with B-F and each of the polynomials.
\begin{table}[h]
\centering
    \begin{tabular}{|c|c|c|c|c|c|}
    \hline
                          & B-F                   & $I_{\boldsymbol{12}}F$  &$Q_1$               &$Q_2$               & $Q_3$ \\ \hline
    $\sigma^2_{t_0}$      & $6.69\cdot 10^{-5}$   & $6.71\cdot 10^{-5}$     &$7.18\cdot 10^{-5}$ &$7.23\cdot 10^{-5}$ &$7.07\cdot 10^{-5}$ \\
    $\beta_0$             & $2.53\cdot 10^{-7}$   & $1.72\cdot 10^{-7}$     &$1.04\cdot 10^{-7}$ &$1.11\cdot 10^{-7}$ &$1.56\cdot 10^{-7}$ \\
    $\beta_1$             & $0.918$               & $0.924$                 &$0.930$             &$0.931$             &$0.930$           \\
    $\beta_2$             & $0.041$               & $0.037$                 &$0.035$             &$0.034$             &$0.035$           \\
    $(\lambda + \theta)$  & $1.014$               & $1.035$                 &$1.020$             &$1.015$             &$1.017$            \\ \hline
\end{tabular}
\caption{\label{Artconsist}Parameter values estimation obtained with $B-F$, $I_{\boldsymbol{12}}F$, $Q_1$, $Q_2$ and $Q_3$.}
\end{table}

The number of contracts in the sample is 264. As we can see in Table \ref{Artconsist}, the numerical errors which raise from the interpolation or the Reduced Basis technique result in different parameter estimations, more observable in the values of $\sigma^2_{t_0}$. Nevertheless, note that  the values obtained for $\beta_1$, $\beta_2$ and $(\lambda + \theta)$ are quite close, resulting probably in very close stochastic processes dynamics (maturities grow up to almost one year). Value $\beta_0$ is quite small, and its influence in the option price might be very small, being a more difficult parameter to estimate exactly.

Now let us check the errors in the estimation. Table \ref{Artconsisterrorin} shows the maximum/mean contract prices and the highest/mean absolute errors committed in the estimation.

\begin{table}[h]\small
\centering
    \begin{tabular}{|c|c|c|c|c|c|c|}
    \hline
                & Price         & Error B-F           & Error $I_{\boldsymbol{12}}F$ & Error $Q_1$          & Error $Q_2$         & Error $Q_3$ \\ \hline
    Max         & $0.2488$      & $6.5\cdot 10^{-4}$ & $12\cdot 10^{-4}$          &$9.5\cdot 10^{-4}$   &$9.6\cdot 10^{-4}$ &$9.4\cdot 10^{-4}$ \\
    Mean        & $0.0689$      & $1.4\cdot 10^{-4}$ & $2.2\cdot 10^{-4}$           &$1.8\cdot 10^{-4}$ &$2.0\cdot 10^{-4}$ &$2.2\cdot 10^{-4}$ \\ \hline
\end{tabular}
\caption{\label{Artconsisterrorin} Maximum/Mean Absolute errors of the estimation with B-F, $I_{\boldsymbol{12}}F$, $Q_1$, $Q_2$ and $Q_3$. }
\end{table}

We remark that there are several different errors in Table \ref{Artconsisterrorin}. The first one, the error labeled $B-F$, is the error of the adequacy of the model. This error appears because we are approximating the continuous SV model with the discrete NGARCH model. Indeed, the literature shows that if we had employed real market data, this error would have been one or two orders of magnitude bigger.

The second error in Table \ref{Artconsisterrorin} is the difference between column $I_{\boldsymbol{12}}F$ with respect to $B-F$, which is the interpolation error. This difference can be made as small as we want just increasing the number of interpolation points (see Section \ref{RBFAOVPi}).

The third error is the difference between columns $Q_1$, $Q_2$, $Q_3$ with respect to $I_{\boldsymbol{12}}F$. This difference comes from the Reduced Basis approach and can be made as small as we want just reducing the value $\epsilon$ (see (\ref{Ch1forhierproc})).

\

Now, let us try to predict future option prices (Out). Suppose that the stock price and maturities have changed. Let $S=0.79,0.81,0.83,...,1.17$ and $T=25,55,85,...,325$ days. We assume that the rest of the market parameter values have not changed. With the parameters estimated in Table \ref{Artconsist} we compute the new option prices with each of the methods.

On the other side, we compute the exact option prices with the SV model and the parameter values $\Omega^{*}$ and compare with the predictions. The errors with respect to the exact option prices are summarized in Table \ref{Artconsisterrorout}.

\begin{table}[h]\small
\centering
    \begin{tabular}{|c|c|c|c|c|c|c|}
    \hline
                & Price         & Error B-F           & Error $I_{\boldsymbol{12}}F$ & Error $Q_1$          & Error $Q_2$         & Error $Q_3$ \\ \hline
    Max         & $0.2367$      & $5.3\cdot 10^{-4}$ & $11\cdot 10^{-4}$          &$9.9\cdot 10^{-4}$   &$10\cdot 10^{-4}$ &$9.5\cdot 10^{-4}$ \\
    Mean        & $0.0669$      & $1.5\cdot 10^{-4}$ & $2.3\cdot 10^{-4}$           &$1.8\cdot 10^{-4}$ &$2.0\cdot 10^{-4}$ &$2.1\cdot 10^{-4}$ \\ \hline
\end{tabular}
\caption{\label{Artconsisterrorout} Maximum/Mean Absolute errors of the prediction with $I_{\boldsymbol{12}}F$, $Q_1$, $Q_2$ and $Q_3$. }
\end{table}

If we compare with Table \ref{Artconsisterrorin}, we can check that the errors committed in the prediction are of the same magnitude with all the numerical methods. We also remark that the experiment that we have realized can be seen as a consistency analysis. Although the parameters obtained with each of the polynomials are slightly different (see Table \ref{Artconsist}), the prices obtained in the prediction are fairly closed to the exact ones. Therefore, the Reduced Basis method can be successfully applied to estimate model parameters/predict option prices. Furthermore,  while we need several minutes to estimate/predict with B-F or $I_{\boldsymbol{12}}F$, we can do the same computations in a few seconds with polynomials $Q_1$, $Q_2$ or $Q_3$.

We remark again that our objective is not to study the approximation of the NGARCH model to the SV model, but to compare the results in the estimation and prediction of the NGARCH model (B-F method) with the results of the interpolation polynomial $I_{\boldsymbol{12}}F$ and the polynomials obtained in the Reduced Basis approximation $Q_1$, $Q_2$ and $Q_3$. The experiment was repeated for several times with different values for $\Omega^{*}$ of the SV model, both picked by the authors or parameters employed/estimated in \cite{HestonSV} and \cite{Papanicolaou}. Obviously, in the estimation we obtained different values for the parameters and the errors were slightly different, but the behaviour between the different errors ($I_{12}$ vs B-F, $Q_1$ vs $I_{12}$,...) remained the same.

\section{Concluding remarks}

In this paper, we have proposed a Chebyshev Reduced Basis Function method in order to deal with the ``Curse of Dimensionality'' which appears when we deal with multidimensional interpolation. The main objective of the work was to reduce the computing time and storing costs which appear in multidimensional models. We have also applied this technique to the practical problem of the real-time option pricing / model calibration problem in financial economics with satisfactory results.

Further work may include a formal comparison between the work presented and other numerical methods employed in option pricing. This is not a straightforward task. First of all (see \cite{Christoffersen}, for example), it is not easy to determine which model (GARCH, SV, Jumps,...) may give the best results for option pricing. Each model employs a different number of dimensions, it may lead to different numerical problems and, probably, several numerical methods have been proposed to approximate option prices.

Other line of work, which we believe it might be of high interest, is to study if the ideas presented in this work could be combined someway with other numerical techniques which can be found in the literature.

\end{document}